\documentclass[twocolumn,floatfix,prl,aps,showpacs]{revtex4}
\usepackage{epsfig,graphicx,tabularx}
\usepackage{subfigure}
\usepackage{amsmath,amsfonts,amssymb}
\usepackage{color,braket}

\newcommand{\be}{\begin{equation}}
\newcommand{\ee}{\end{equation}}

\begin{document}

\title{Quantum filaments in dipolar Bose-Einstein condensates}
\author{F. W\"achtler}
\author{L. Santos}
\affiliation{Institut f\"ur Theoretische Physik, Leibniz Universit\"at Hannover, Appelstr. 2, DE-30167 Hannover, Germany}

\begin{abstract}
Collapse in dipolar Bose-Einstein condensates may be arrested by quantum fluctuations.  
Due to the anisotropy of the dipole-dipole interactions, 
the dipole-driven collapse induced by soft excitations is compensated by the repulsive Lee-Huang-Yang contribution resulting from quantum fluctuations 
of hard excitations, in a similar mechanism as that recently proposed for Bose-Bose mixtures. 
The arrested collapse results in self-bound filament-like droplets, 
providing an explanation to recent dysprosium experiments. Arrested instability and droplet formation are
novel general features directly linked to the nature of the dipole-dipole interactions, and should hence 
play an important role in all future experiments with strongly dipolar gases.

\end{abstract}


\maketitle



Dipole-dipole interactions~(DDI) lead to qualitatively new physics for dipolar gases compared to non-dipolar ones~\cite{Lahaye2009,Baranov2012}. 
As a result, this physics constitute the focus of a large interest, including experiments on magnetic atoms~\cite{Griesmaier2005,Lu2011,Aikawa2012,DePaz2013}, 
polar molecules~\cite{Ni2008,Yan2013,Takehoshi2014,Park2015}, and Rydberg-dressed atoms~\cite{Rydberg-Dressed}. A characteristic feature of dipolar Bose-Einstein condensates~(BECs) is their geometry-dependent 
stability~\cite{Koch2008}. If the condensate is elongated along the dipole orientation, the DDI are attractive in average, 
and the BEC may become unstable, in a similar, but not identical, way as a BEC with negative $s$-wave scattering length, $a<0$. 
Chromium experiments showed that, as for $a<0$, the unstable BEC collapses, albeit with a peculiar $d$-wave post-collapse dynamics~\cite{Lahaye2008}.

This picture has been challenged by recent dysprosium experiments~\cite{Kadau2015}, in which destabilization, induced by a quench to a sufficiently low $a$, is not followed by collapse, but rather by the formation of 
stable droplets that are only destroyed in a large time scale by weak three-body losses~(3BL). 
This surprising result, which resembles the Rosensweig instability in ferrofluids~\cite{Cowley1967,Timonen2013}, points to an up to now unknown stabilization mechanism that 
plays a similar role as that of surface tension in classical ferrofluids. It has been recently suggested that large conservative three-body forces, with a strength 
several orders of magnitude larger than the 3BL, may account for the observation~\cite{Xi2015,Bisset2015}. There is however no justification of why large 
three-body forces should be present, or whether there is a link between them and the DDI.

\begin{figure}[t]
\begin{center}
\includegraphics[width=\columnwidth]{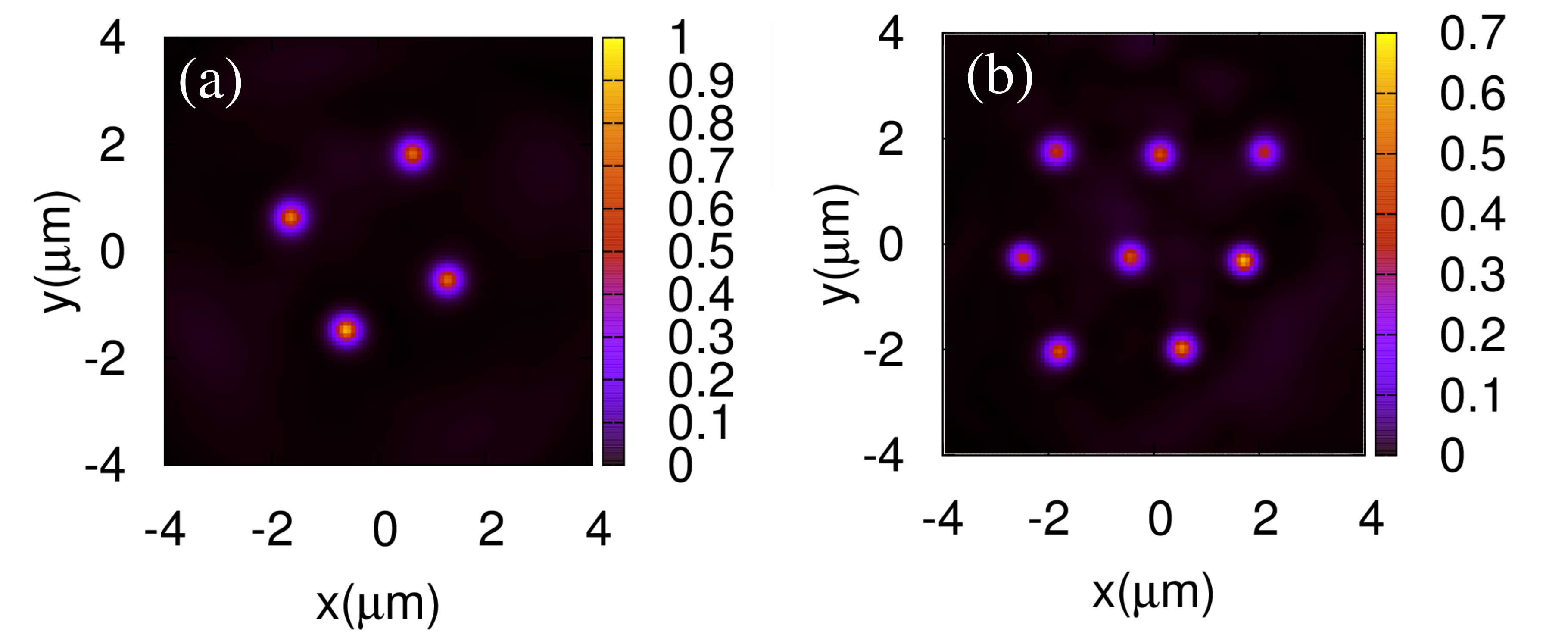}
\end{center}
\vspace*{-0.4cm}
\caption{(Color online) Crystal-like droplet arrangements of $n_{XY}(x,y)/N$, with $n_{XY}(x,y)=\int dz\, n({\mathbf r})$, for a BEC of N=7500 atoms~(top), and 15000 atoms~(bottom), 
initially formed with $a=120a_B$, $20$ms after a quench to $a=70a_B$.}
\label{fig:1}
\end{figure}

This Letter explores an alternative mechanism, based on quantum fluctuations, which is suggested by very recent experiments~\cite{Pfau-Private}. 
As recently shown~\cite{Petrov2015}, Lee-Huang-Yang~(LHY) corrections may stabilize 
droplets in unstable Bose-Bose mixtures. This interesting effect results from the presence of soft and hard elementary excitations. Whereas soft modes may become 
unstable, quantum fluctuations of the hard modes may balance the instability, resulting in an equilibrium droplet. As shown below,  due to the anisotropy of the DDI, 
a dipolar BEC also presents soft and hard modes, characterized in free space by momenta perpendicular or parallel to the dipole orientation. 
As a result, the LHY correction resulting from the hard modes provides a repulsive term that dominates at large densities arresting local collapses, resulting in 
the nucleation of droplets~(see Figs.~\ref{fig:1}). We show by means of a generalized nonlocal non-linear Schr\"odinger equation~(NLNLSE)  that this mechanism 
accounts for the Dy experiments. We stress that this effect results from the peculiar nature of the DDI, being hence a characteristic novel feature of strongly dipolar gases, which 
should play an important role in future experiments with highly magnetic atoms and polar molecules.

\paragraph{Generalized NLNLSE--}  We consider a BEC of magnetic dipoles of mass m and dipole moment ${\boldsymbol \mu}$ oriented along the 
$z$ direction by an external magnetic field~(equivalent results can be found for electric dipoles). In mean-field~(MF), the physics is given by the NLNLSE~\cite{Lahaye2009}:
\begin{equation}
i \hbar \dot \psi({\mathbf r})=\left [ \frac{-\hbar^2\nabla^2}{2m}+V({\mathbf r})+g|\psi({\mathbf r})|^2+\Phi({\mathbf r}) \right ] \psi({\mathbf r}),
\label{eq:NLNLSE}
\end{equation}
with $\psi({\mathbf r})$ the BEC wavefunction, $V({\mathbf r})$ the trapping potential, $g=\frac{4\pi\hbar^2 a}{m}$, and 
$\Phi({\mathbf r})=\int d^3r' V_{dd}({{\mathbf r}-{\mathbf r}')|\psi(\mathbf r'})|^2$, with $V_{dd}({\mathbf r})=\frac{\mu_0 |{\boldsymbol \mu}|^2}{4\pi r^3}(1-3\cos^2\theta)$, where $\mu_0$ 
is the vacuum permittivity, and $\theta$ is the angle between ${\mathbf r}$ and ${\boldsymbol\mu}$. 


In the homogeneous case, $V({\mathbf r})=0$ with density $n$, elementary excitations with momentum ${\mathbf k}$ have an energy 
$E({\mathbf k})=\sqrt {\epsilon_k \left (\epsilon_k +2gn f(\epsilon_{dd},\theta_k) \right ) }$, 
where $\epsilon_k=\frac{\hbar^2k^2}{2m}$, and $f(\epsilon_{dd},\theta_k)=1+\epsilon_{dd}(3\cos^2\theta_k-1)$, with  
$\epsilon_{dd}=\frac{\mu_0|{\boldsymbol \mu}|^2}{3g}$, and $\theta_k$ the angle between ${\mathbf k}$ and ${\boldsymbol\mu}$. 
Due to the anisotropy of the DDI, excitations with $\cos^2\theta_k>1/3$ 
become harder with growing $\epsilon_{dd}$, whereas those with $\cos^2\theta_k<1/3$ become softer. 
For $\epsilon_{dd}>1$, long wave-length excitations with $\theta_k=\pi/2$ drive the BEC unstable.
Quantum fluctuations of the excitations result in the LHY correction of the chemical potential~\cite{Lima2011,Lima2012}:
\begin{equation}
\Delta \mu (n,\epsilon_{dd})=\frac{32}{3\sqrt{\pi}}gn\sqrt{na^3}F(\epsilon_{dd}),
\end{equation}
with $F(\epsilon_{dd})=\frac{1}{2}\int d\theta_k \sin{\theta_k} f(\epsilon_{dd},\theta_k)^{5/2}$.
In the vicinity of the instability, $\epsilon_{dd}\sim 1$, the overwhelming contribution to $F(\epsilon_{dd})$ stems from hard modes~($\cos^2\theta_k>1/3$). 
Crucially, this is true even when the BEC becomes unstable. This situation, with unstable soft modes and LHY correction dominated by stable hard modes, resembles the 
recently discussed case of Bose-Bose mixtures~\cite{Petrov2015}. As for that scenario,  
the contribution of the unstable soft modes is negligible for 
$\epsilon_{dd}\sim 1$, and quantum fluctuations of the hard modes result in a repulsive LHY correction $\propto n^{3/2}$.


Let us consider at this point a harmonically trapped BEC,  $V({\mathbf r})=\frac{1}{2}m(\omega_x^2 x^2+\omega_y^2 y^2+\omega_z^2 z^2)$. 
The treatment of beyond MF corrections is in general much more involved. In the Thomas-Fermi~(TF) regime one may evaluate the effect of quantum 
fluctuations by treating the excitations quasi-classically and employing local density approximation~(LDA), obtaining a corrected equation of state~\cite{Lima2011,Lima2012}: 
$\mu(n({\mathbf r}))=V({\mathbf r})+\mu_{0}(n({\mathbf r}),\epsilon_{dd})+\Delta \mu (n({\mathbf r}),\epsilon_{dd})$, with 
$\mu_{0}(n({\mathbf r}),\epsilon_{dd})=gn({\mathbf r})+\int d^3 r' V_{dd}({\mathbf r}-{\mathbf r}')n({\mathbf r}')$. 
One may then insert this correction in a generalized NLNLSE:
\begin{equation}
 i \hbar \dot \psi({\mathbf r})\! =\!\left [\hat H_0 \! +\! \mu_{0}(n({\mathbf r}),\epsilon_{dd})\! +\! \Delta \mu (n({\mathbf r}),\epsilon_{dd}) \right ]\! \psi({\mathbf r}), 
\label{eq:gNLNLSE}
\end{equation}
with $\hat H_0\equiv \frac{-\hbar^2\nabla^2}{2m}+V({\mathbf r})$.  
This equation is appealing since it allows for a simplified analysis of the effects of quantum fluctuations in the TF regime, 
and because it may be simulated using the same numerical techniques  
employed for Eq.~\eqref{eq:NLNLSE}~\cite{Lahaye2009,footnote-numerics}.
However the use of the LDA to evaluate the effects of quantum fluctuations in quench experiments must be carefully considered. The droplets discussed below are in the TF regime 
along the dipole direction in all cases, whereas only large droplets are in the TF regime also along $xy$. 
For small droplets, with less than $4000$ atoms in the calculations below, the $xy$ density profile approaches rather a Gaussian.  
We may evaluate the contribution of quasi-classical excitations with momenta $\mathbf{k}$, such that 
$|\mathbf{k}| R(\theta_k)\gg 1$ where $R(\theta_k)$ is a typical distance for density variation in the droplet along the direction given by the angle $\theta_k$. 
This contribution is for the smallest droplets presented below of the order of $\sim 80\%$ of the total LHY correction expected from LDA~(for details of this estimation see \cite{footnote-cutoff}). 
The correction due to long wave-length modes 
may hence modify the prefactor of the correction, but the bulk of the effect is well recovered by Eq.~\eqref{eq:gNLNLSE}. 
We postpone for a future analysis the detailed study of the effect of long wavelength excitations. 
In addition, the validity of the generalized NLNLSE demands a small quantum depletion~\cite{Lima2011,Lima2012}, 
$\eta({\mathbf r})\equiv \frac{\Delta n({\mathbf r})}{n({\mathbf r})}=\frac{8}{3\sqrt{\pi}}\sqrt{n({\mathbf r})a^3}F_D(\epsilon_{dd})$, 
with $F_D(\epsilon_{dd})=\frac{1}{2}\int d\theta_k \sin{\theta_k} f(\epsilon_{dd},\theta_k)^{3/2}$.  
In our simulations, $\eta({\mathbf r})\lesssim 0.01$ at any point and time.


\paragraph{Droplet nucleation--} 
In the following we employ Eq.~\eqref{eq:gNLNLSE} to study the formation of BEC droplets in recent Dy experiments~\cite{Kadau2015}. 
We consider a BEC with $N$ Dy atoms, with $|{\boldsymbol \mu}|=10\mu_B$, with $\mu_B$ the Bohr magneton. In order to compare our results with recent 
experiments we assume  a trap with
$\omega_{x,y,z}/2\pi=(44,46,133)$Hz~\cite{footnote-trap}. 
We employ imaginary time evolution of Eq.~\eqref{eq:gNLNLSE} to form an initial BEC with $a=120a_B$, with $a_B$ the Bohr radius. Under these conditions the BEC, with a wavefunction $\psi_0({\mathbf r})$, is stable and in the TF regime. 
At finite temperature, $T$, thermal fluctuations seed the modulational instability after the quench of $a$ discussed below, and may hence influence droplet nucleation. 
Following Ref.~\cite{Bisset2015} we add thermal fluctuations~(for $T=20$nK) in the form $\psi({\mathbf r},t=0)=\psi_0({\mathbf r})+\sum_n\alpha_n\phi_n$, where $\phi_n$ are eigenmodes of the 
harmonic trap with eigenenergies $\epsilon_n$, the sum is restricted to $\epsilon_n<2k_B T$, and $\alpha_n$ is a complex Gaussian random variable 
with $\langle |\alpha_n|^2\rangle= \frac{1}{2}+(e^{\epsilon_n/k_B T}-1)^{-1}$~\cite{footnote-SGPE}.

At $t=0$ we perform a quench in $0.5$ms to a final $a=70 a_B$ that 
destabilizes the BEC~\cite{footnote-a}. The most unstable Bogoliubov mode has a non-zero angular momentum, a so-called angular roton~\cite{Ronen2007}, and  
as a result at $T=0$ the BEC develops an initial ring-like modulational instability on the $xy$ plane, followed by azimuthal symmetry breaking into droplets. 
At finite $T$ droplets may nucleate from thermal fluctuations 
before the ring-like structure associated with the angular-roton instability develops~(as it is observed in experiments~\cite{Pfau-Private-2}). 
Both cases are characterized by the formation of stable droplets in few ms, which eventually arrange in a quasi-crystalline structure as those of Figs.~\ref{fig:1}, in excellent agreement with 
the experimental results of Ref.~\cite{Kadau2015}. 
Droplet nucleation does not involve however the whole condensate. 
A significant amount of atoms remains in a halo-like background too dilute to gather particles into a stable droplet~(approximately $30\%$ in Figs.~\ref{fig:1}, although it is barely visible 
due to the contrast).


\paragraph{Droplet features --} The droplets result from the compensation of the attractive MF term $\mu_0\propto n({\mathbf r})$ by
the effective repulsion introduced by the LHY term, $\Delta\mu\propto n({\mathbf r})^{3/2}$. In order to study the properties of individual droplets, we 
evolve Eq.~\eqref{eq:gNLNLSE} in imaginary time for $a=70a_B$ and different particle numbers. In order to guarantee the controlled formation of a single droplet in the numerics, we employ as initial condition 
for the imaginary time evolution a cigar-like Gaussian wavefunction at the trap center very compressed on the $xy$ plane~\cite{footnote-initial} . 

\begin{figure}[t]
\begin{center}
\includegraphics[width=\columnwidth]{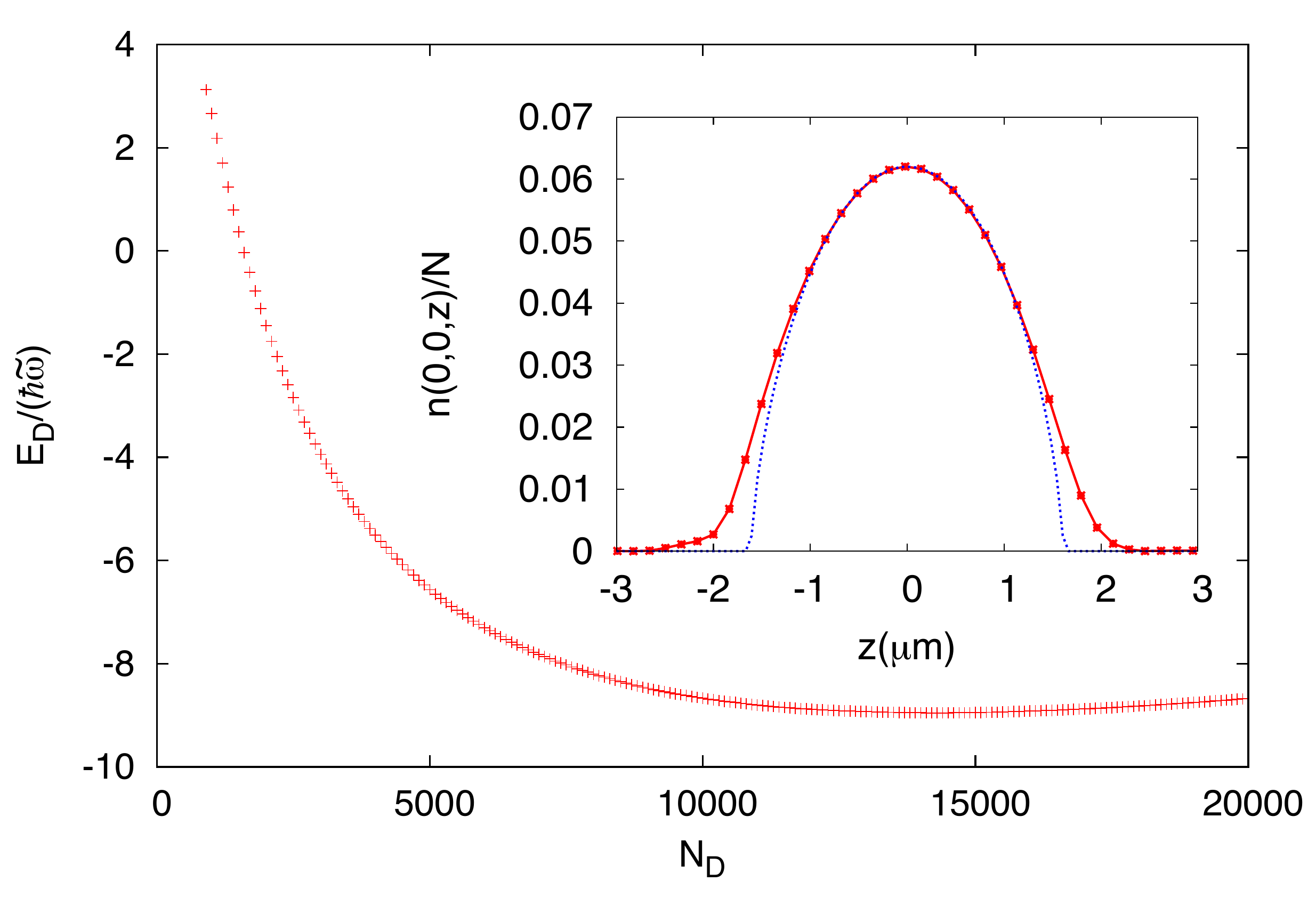}
\end{center}
\vspace*{-0.2cm}
\caption{(Color online) Droplet energy, $E_D$~(in units of $\hbar\tilde\omega$, with $\tilde\omega=(\omega_x\omega_y\omega_z)^{1/3}$) as a function of the number of particles in the droplet, $N_D$, 
for $a=70a_B$; for $N_D<N_{min}\simeq 900$ no stable droplet is found. Droplets with positive internal energy occur for $N_D\lesssim 1500$. In the inset, we show the density profile (solid line with crosses) 
of a droplet with $N_D=1000$ at the trap center for the cut $x=y=0$. At the center of the droplet,  $n(0,0,z)\propto (1-z^2/Z^2)^{2/3}$~(dotted curve).}
\label{fig:2}
\end{figure}

Figure~\ref{fig:2} shows the droplet energy, $E_D$, as a function of the number of particles in the droplet, $N_D$. Two important features are worth mentioning. 
There is a minimal particle number, $N_{min}\simeq 900$, such that for $N_D<N_{min}$ 
no stable droplet may form.  If the local density does not allow for the gathering of that critical number, then no 
droplet is formed, accounting for the background halo. Second, $E_D(N_D)$ presents a non-monotonous dependence with $N$, showing a minimum~(at $N_D\simeq 13000$ in Fig.~\ref{fig:2}), being only 
positive at $N_D$ values close to $N_{min}$~($E_D=0$ at $N_D\simeq 1500$ in Fig.~\ref{fig:2}). This is particularly relevant for the droplet nucleation after a quench. 
After the quench, the BEC energy, which is initially positive~\cite{footnote-Energy},  is almost conserved, just decreasing slowly due to 3BL. The final droplet gas is characterized by the internal energy of the droplets, the center of mass~(CM), kinetic and potential, energy of the droplets, and the inter-droplet dipole-dipole repulsion (the 
halo, being much more dilute has a comparatively small contribution to the BEC energy). Although the CM energy and the repulsive inter-droplet interaction are obviously positive, they 
cannot balance a negative internal energy of the droplets, as required by the quasi-conservation of the energy  in the absence of strong dissipation. This explains why, as discussed below, in the quench experiments droplets form with particle numbers between $900$ and $1500$, despite the fact that bigger droplets could be in principle stable~(Fig.~\ref{fig:2}).

The shape of the droplets is also significant for the overall discussion. 
The droplets are markedly elongated along the dipole direction. For $N_D\simeq 1000$, the $z$ half-size is $\simeq 2\mu$m, whereas along $xy$ is $\lesssim 0.3\mu$m. 
This is expected, since a cigar-like shape is required for an attractive DDI that overwhelms the repulsive 
contact MF term. In addition, as mentioned above, the droplets are 
in any case well within the TF regime along the $z$ direction~(see the inset of Fig.~\ref{fig:2} for $N_D=1000$). Large droplets, with $N_D>8000$, are also well within the TF regime along the $xy$ direction. 
On the contrary as already noted, small droplets with $N_D<4000$ have approximately a Gaussian profile along $xy$. 
This remains true for the droplets found in the simulations of quench experiments. 
Quantum pressure is hence non-negligible for small droplets, but it is not crucial for the droplet stability, which is provided by the compensation of the attractive MF interaction and the LHY correction.
This is in stark contrast with self-bound solitons that require necessarily quantum pressure to compensate attractive interactions, and hence cannot occur in any case in the TF regime. 
Due to the LHY term, the droplets do not present an inverted-paraboloid profile even in the TF regime.  At the center,
$n(x=0,y=0,z)\simeq (1-z^2/Z^2)^{2/3}$~(inset of Fig.~\ref{fig:2}). 


\begin{figure}[t]
\begin{center}
\includegraphics[width=\columnwidth]{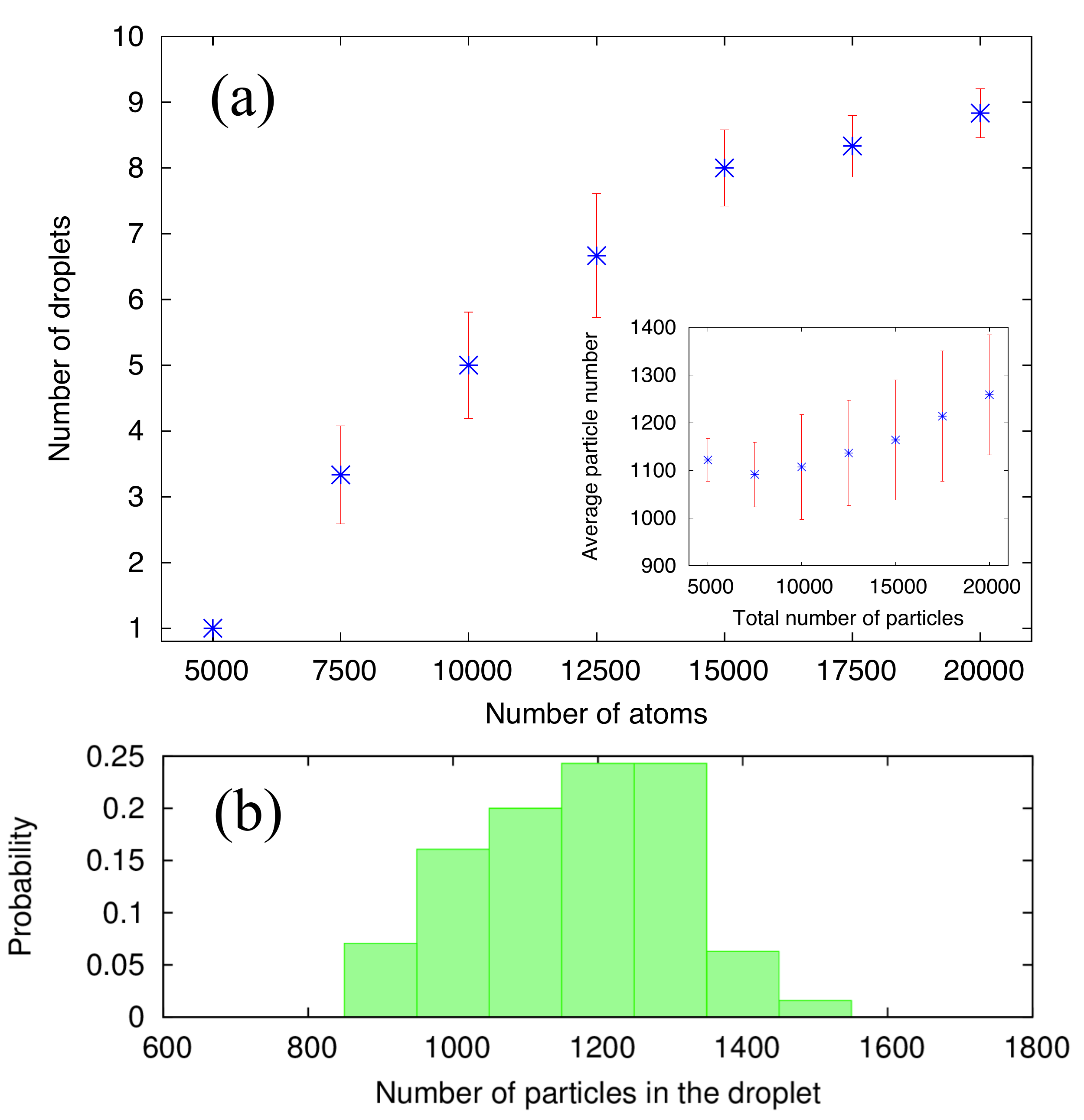}
\end{center}
\vspace*{-0.2cm}
\caption{(Color online) (a) Number of droplets~(see text) as a function of the initial number of atoms $20$ms after a quench from $a=120 a_B$ to $70a_B$; the inset shows the 
number of particles per droplet as a function of $N$ under the same conditions. In both cases the average value is denoted by a blue cross, and the variance by the error bar; 
(b) Histogram of the number of particles in a droplet again for the same conditions. 
The histogram was evaluated from a sample of $260$ droplets. }
\label{fig:3}
\end{figure}

\paragraph{Droplet statistics in quench experiments--} We have performed for different $N$  
simulations of the BEC dynamics after the quench of $a$, starting from different initial conditions given by random thermal fluctuations.
As in the experiments, we observe that the droplets arrange in crystal-like patterns~(see Fig.~\ref{fig:1}), although they present a residual dynamics.
Our simulations show that  the number of particles in a droplet varies from droplet to droplet in a single shot and between shots (as in the actual experiments). 
This variance results from the fact that stable droplets, as shown above, may be formed for different values $N_D>N_{min}$. 
The corresponding variance of the number of droplets is additionally affected by  the variable importance of the background halo. 
In addition, droplets formed at the verge of instability $N_D\simeq N_{min}$, may eventually become unstable and dilute in the halo, and hence the number of apparent droplets may vary in time. In order to 
measure objectively the number of droplets, we have obtained the column density $n_{XY}(x,y)\equiv\int dz\, n({\mathbf r})$ after $20$ ms of post-instability dynamics, and defined a droplet as such if it reaches a maximal 
$n_{X,Y}/N>0.3$~(see e.g. Figs.~\ref{fig:1}). Figure~\ref{fig:3}(a) summarizes our results for the dependence with $N$ of the number of droplets formed after $20$ ms of post-instability dynamics. Although as mentioned 
above the number of droplets presents a relevant statistical variance, the average number shows in agreement with experiments an approximate linear dependence.
The deviation at larger $N$ is due to the fact that at $20$ms there are a number of droplets that are about to be nucleated in the outer halo regions but are not fully formed (according to the previous criterion). 
The deviation at low $N$ is due to the longer time needed to fully develop droplets (e.g. $N=5000$ develops up to $3$ droplets after $40$ms). 
The approximate linear dependence of Fig.~\ref{fig:3}(a) stems from the local character of the nucleation, which results in a number of particles per droplet that is basically 
independent of $N$~(inset of Fig.~\ref{fig:3}(a)). 
The histogram of Fig.~\ref{fig:3}(b) shows that, as expected from our discussion of the droplet energy, the particle number per droplet lies overwhelmingly between $900$ and $1500$, 
with an average of approximately $1200$, again in good agreement with experiments.

\begin{figure}[t]
\begin{center}
\includegraphics[width=\columnwidth]{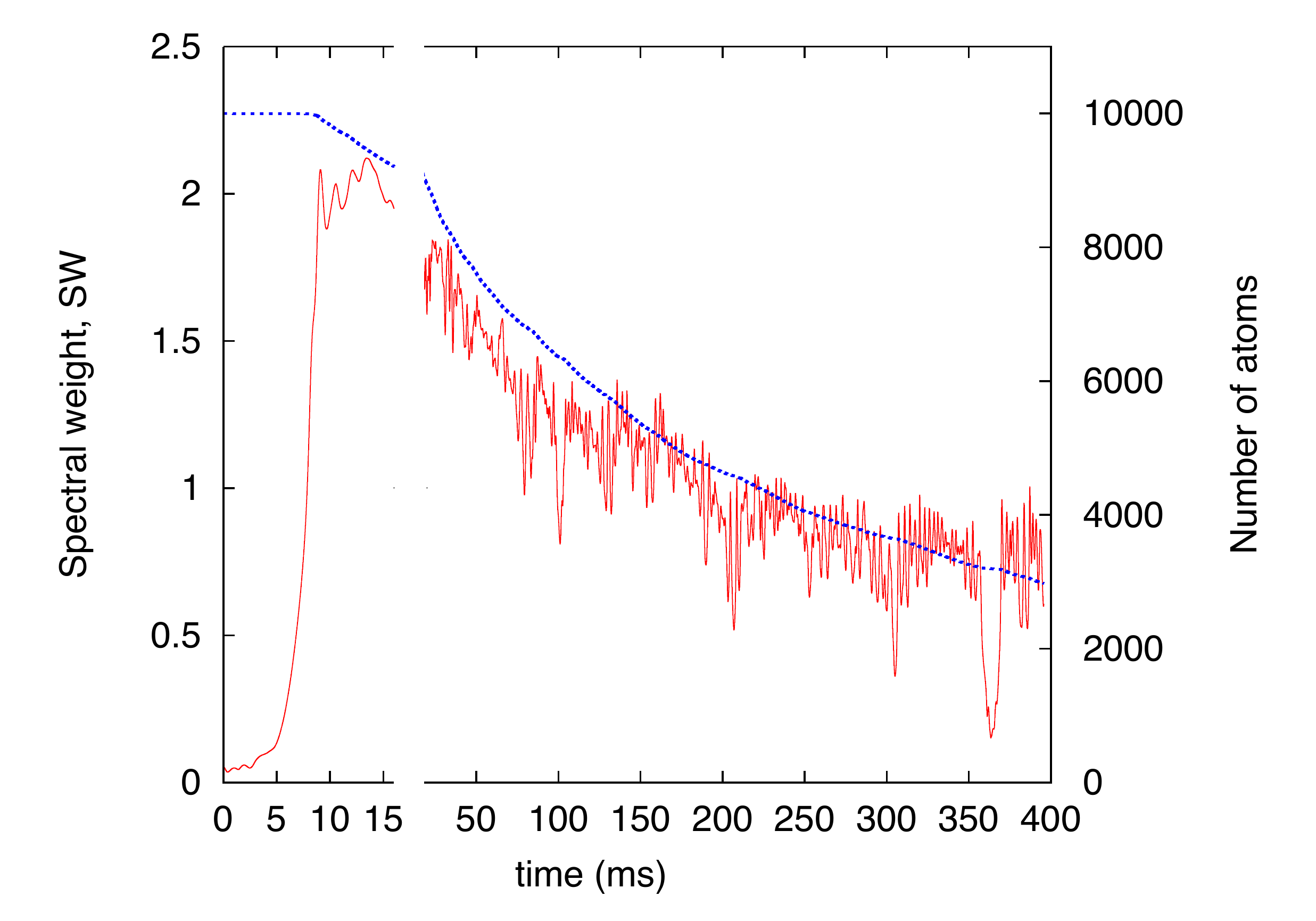}
\end{center}
\vspace*{-0.2cm}
\caption{(Color online) Number of atoms~(blue dotted) and spectral weight~(red solid)  as a function of the time after a quench from $a=120 a_B$ to $70a_B$ for a BEC with initially $N=10000$ atoms.}
\label{fig:4}
\end{figure}

\paragraph{Three-body losses--} 
We observe in our Dy simulations peak densities of $\sim 2\times 10^{21}$m$^{-3}$~\cite{footnote-peak}. At these densities, albeit low, 3BL 
become relevant in the long run. In order to take them into account, we add a term $-i\frac{\hbar L_3}{2}|\psi({\mathbf r})|^4 \psi({\mathbf r})$~\cite{Lahaye2008} to the right hand side of Eq.~\eqref{eq:gNLNLSE}, with 
$L_3=1.2\times 10^{-41}$m$^{6}/$s~\cite{footnote-L3}. 
Figure~\ref{fig:4} shows for $N=10000$ the number of atoms as a function of time. We show in the same figure, the spectral weight 
$SW=\int d^2k \tilde n_{XY}({\mathbf k})$, where $\tilde n_{XY}({\mathbf k})$ is the Fourier transform of the column density $n_{XY}(x,y)\equiv \int dz\, n({\mathbf r})$ and 
the momentum integral extends from $k_{min}=1.5\mu$m$^{-1}$ to $k_{max}=5\mu$m$^{-1}$. 
This function was introduced in Ref.~\cite{Kadau2015} to characterize the appearance and disappearance of the droplet pattern.  
The losses not only decrease the atom number, but also  lead to the eventual destruction of the droplets, which may loose too much particles to remain stable against 
melting in the background. Moreover, 3BL eliminates high-energy atoms at the BEC maxima. This energy dissipation 
plays an important role in the eventual nucleation of further droplets, and especially in the formation of crystal-like droplet arrangements as those of Fig.~\ref{fig:1}, which minimize inter-droplet interaction. 
The results of Fig.~\ref{fig:4} are in very good agreement with experiments, showing a growth of SW up to $t\sim 10$ms, and a subsequent decrease in a much longer time scale, accompanied by 
the corresponding particle loss due to 3BL.

\paragraph{Conclusions and outlook--} Quantum fluctuations prevent local collapses in unstable dipolar BECs. 
In particular, our results show the appearance of filament-like droplets, accounting for recent results in Dy condensates.
Since the LHY correction depends on $na^3$, we expect that droplets should collapse for lower $a$ values, providing a clear criterion to discern LHY stabilization from 
stabilization based on large three-body forces~\cite{Xi2015,Bisset2015}. Our results, based on a simplified treatment using a generalized NLNLSE are already in very 
good agreement with the experiments~\cite{Kadau2015,Pfau-Private}, although a 
more precise analysis of the effects of long-wave length excitations in small droplets may be necessary to provide a fully quantitative comparison, in particular in what concerns 
the peak density in the droplets. We stress that LHY stabilization results from the anisotropy of the dipolar interactions. It was 
absent in previous Cr experiments~\cite{Lahaye2008} because the BEC became unstable for a value of $a\simeq $ 
$10$ times smaller than in Dy ($m|{\boldsymbol \mu}|$ in Dy is $10$ times larger than in Cr), and LHY stabilization would demand $n\sim 10^{24}$m$^{-3}$, a density 
which is never reached due to 3BL. In contrast, LHY stabilization and droplet nucleation are a characteristic general feature induced by the DDI that may play a role in all future 
experiments with strongly dipolar gases of highly-magnetic atoms and polar molecules.



{\it Acknowledgements.--} We thank D. Petrov and Y. Li for interesting comments, and 
H. Kadau, I. Ferrier-Barbut, and T. Pfau for insightful discussions and for providing us with their experimental results.
 We acknowledge support by the cluster QUEST, and the DFG Research Training Group 1729.




\thebibliography{99}

\bibitem{Lahaye2009} See e.g. T. Lahaye  {\it et al.}, Rep. Prog. Phys. {\bf 72}, 126401 (2009), and references therein.

\bibitem{Baranov2012} See e.g.  M. A. Baranov, M. Dalmonte,  G. Pupillo, and P. Zoller, Chem. Rev. {\bf 112}, 5012 (2012), and references therein.

\bibitem{Griesmaier2005} A. Griesmaier  {\it et al.}, Phys. Rev. Lett. {\bf 94}, 160401 (2005).

\bibitem{Lu2011} M. Lu, N. Q. Burdick, S. H. Youn, and B. L. Lev, Phys. Rev. Lett. {\bf 107}, 190401 (2011).

\bibitem{Aikawa2012} K. Aikawa  {\it et al.}, Phys. Rev. Lett. {\bf 108}, 210401 (2012).

\bibitem{DePaz2013} A. de Paz {\it et al.}, Phys. Rev. Lett. {\bf 111}, 185305 (2013).

\bibitem{Ni2008} K. K. Ni {\it et al.}, Science {\bf 322}, 231 (2008).
\bibitem{Yan2013} B. Yan  {\it et al.}, Nature {\bf 501}, 521 (2013).
\bibitem{Takehoshi2014} T. Takekoshi {\it et al.}, Phys. Rev. Lett. {\bf 113}, 205301 (2014).
\bibitem{Park2015} J. W. Park, S. A. Will, and M. W. Zwierlein, Phys. Rev. Lett. 114, 205302 (2015).

\bibitem{Rydberg-Dressed} See e.g. J. B. Balewski {\it et al.}, New J. Phys. {\bf 16}, 063012 (2014).

\bibitem{Koch2008} T. Koch {\it et al.}, Nature Physics {\bf 4}, 218 (2008).
\bibitem{Lahaye2008} T. Lahaye {\it et al.}, Phys. Rev. Lett., {\bf 101}, 080401 (2008).

\bibitem{Kadau2015} H. Kadau {\it et al.}, arXiv:1508.05007.

\bibitem{Cowley1967} M. D. Cowley and R. E.  Rosensweig, Journal of Fluid Mechanics {\bf 30}, 671 (1967).
\bibitem{Timonen2013} J. V. I.  Timonen {\it et al.}, Science {\bf 341}, 253 (2013).

\bibitem{Xi2015} K.-T. Xi and H. Saito, arXiv:1510.07842.
\bibitem{Bisset2015} R. N. Bisset and P. B. Blakie, Phys. Rev. A {\bf 92}, 061603(R) (2015).

\bibitem{Pfau-Private} I. Ferrier-Barbut {\it et al.}, arXiv:1601.03318

\bibitem{Petrov2015} D. S. Petrov, Phys. Rev. Lett. {\bf 115}, 155302 (2015).

\bibitem{Lima2011} A. R. P. Lima and A. Pelster, Phys. Rev. A {\bf 84}, 041604(R) (2011).
\bibitem{Lima2012} A. R. P. Lima and A. Pelster, Phys. Rev. A {\bf 86}, 063609 (2012).

\bibitem{footnote-numerics} The simulation of Eq.~\eqref{eq:gNLNLSE} is performed using split operator techniques, and treating the DDI using convolution theorem and fast-Fourier transformation. 
Following [S. Ronen, D. C. E. Bortolotti, and J. L. Bohn, Phys. Rev. A {\bf 74}, 013623 (2006)] we employ a cut-off of the dipole-dipole potential to reduce spurious boundary effects.

\bibitem{footnote-cutoff} The relative importance of short- and long-wave length excitations in the LHY correction may be estimated as follows.
We consider a low-momentum cut-off $q_c(\theta)=q_z (\cos^2\theta+\lambda^2 \sin^2\theta)^{1/2}$, with $\lambda$ the aspect ratio of the droplet, and $q_z $ the cut-off along $z$. 
Introducing this cut-off in the LHY calculation for the homogeneous space at a given density $n$, results in a modified correction
$
\frac{\Delta\mu_c}{\Delta\mu}=\frac{15\sqrt{2}}{16}\frac{\int_0^\pi d\theta \sin\theta \chi(\epsilon_{dd},\theta)}{\int_0^\pi d\theta \sin\theta f(\epsilon_{dd},\theta)^{5/2}}
$ 
where
$
\chi(\epsilon_{dd},\theta)=2^{5/2}\left ( \frac{ 2 f (\epsilon_{dd},\theta)}{15}-\frac{q_c(\theta)^2}{2} \right ) \left  (  \frac{q_c(\theta)^2}{2}+f(\epsilon_{dd},\theta)   \right )^{3/2}+
\frac{q_c(\theta)^5}{5}+\frac{q_c(\theta)^3}{3}f(\epsilon_{dd},\theta)-\frac{q_c(\theta)}{2}f(\epsilon_{dd},\theta)^2
$, and $q_c(\theta)$ is in units of $\xi^{-1}$, with $\xi=(8\pi n a)^{-1/2}$. A droplet with $N_D=1000$ particles has for $a=70a_B$ a $z$ size of $\simeq 2\mu{\mathrm m}\simeq 25\xi$ (with $\xi$ calculated for an averaged central density of $1.5\times 10^{21}$m$^{-3}$), and an aspect ratio $\lambda\simeq 6$. 
For a $z$ cut-off $q_z\xi \simeq 0.25$, excitations with $|{\mathbf q}(\theta)|>q_c(\theta)$ may be considered as quasi-classical. For this cut-off, we obtain $\frac{\Delta\mu_c}{\Delta\mu}\simeq 0.8$, showing a large 
contribution of quasi-classical excitations to the LHY correction.

\bibitem{footnote-trap} Droplets may form in other trap geometries as well, but 
the details of the stability threshold, as well as of the droplet nucleation vary with the precise trap.

\bibitem{footnote-SGPE} We obtain similar results using a stochastic Gross-Pitaevskii equation to create the thermal excitations.

\bibitem{footnote-a} Quenches to $a=80a_B$ also drive instability for $N\gtrsim 10000$, but  do not destabilize smaller condensates. 

\bibitem{Ronen2007} S. Ronen, D. C. E. Bortolotti, and J. L. Bohn, Phys. Rev. Lett. {\bf 98}, 030406 (2007).

\bibitem{Pfau-Private-2} H. Kadau and T. Pfau, private communication.

\bibitem{footnote-initial} Using other initial conditions, in particular a pancake wavefunction elongated 
on the $xy$ plane,  results in the formation of variable droplet configurations similar as those discussed below in the real time evolution. In passing, this shows that droplet nucleation 
and the formation of (metastable) droplet structures should occur not only in the post-quench dynamics, but also when directly forming the condensates at sufficiently low scattering lengths, as discussed in Ref.~\cite{Kadau2015}. 

\bibitem{footnote-Energy} Of the order of $8\hbar\tilde\omega$ in Figs.~\ref{fig:1}, with $\tilde\omega\equiv(\omega_x\omega_y\omega_z)^{1/3}$. 

\bibitem{footnote-peak} These densities are larger than those in recent experiments~\cite{Pfau-Private} by a factor of $3$--$4$. Note that the LHY depends on $na^3$ and hence a larger $a$ 
may reduce the density. Even more relevantly, as mentioned in the main text, the prefactor of the LHY may be modified for small droplets by short-wave length excitations. An increase in the pre-factor 
will further decrease the peak density.  



\bibitem{footnote-L3} The exact value of $L_3$ is not yet known but it should be in the lower $10^{-41}$m$^{6}/$s. See M. Lu, {\it Quantum Bose and Fermi gases of dysprosium: production and 
initial study}, PhD Thesis, Stanford University (2014).
\end{document}